%% file: sig-alternate-sample.tex
\documentclass{sig-alternate-05-2015}

\usepackage{times}
\usepackage{epsfig}
\usepackage{graphicx}
\usepackage{amsmath}
\usepackage{amssymb}
\usepackage{booktabs}
\usepackage{graphicx}
\usepackage{multirow}
\usepackage{adjustbox}
\usepackage{subcaption}
\usepackage{pgfplots}
\usepackage{tikz}
\usepackage{enumitem}
\usepackage{booktabs}
\usepackage{multirow}
\usepackage[above]{placeins}
\usepackage{amsmath}
\usepackage{amsfonts}
\usepackage{amssymb}
\usepackage{color}
\usepackage[hidelinks]{hyperref}   
\usepackage{adjustbox}
\usepackage{float}
\usepackage{subcaption}
\usepackage{array}
\usepackage{eurosym}
\usepackage{background}
\usepackage{fancyhdr}
\usepackage{transparent}
\usepackage{indentfirst}
\usepackage{etoolbox}
\usepackage{pdfpages}
\usepackage{bbm}
\usepackage{tikz}
\usepackage{enumitem}
\usepackage{lscape}
\usepackage{booktabs}
\usepackage{multirow}

\begin{document}

\CopyrightYear{2016} 
\setcopyright{acmcopyright}
\conferenceinfo{LTA'16,}{October 16 2016, Amsterdam, Netherlands}
\isbn{978-1-4503-4517-0/16/10}\acmPrice{\$15.00}
\doi{http://dx.doi.org/10.1145/2983576.2983582}

\clubpenalty=10000 
\widowpenalty = 10000


\title{Where is my Phone ?}

\subtitle{Personal Object Retrieval from Egocentric Images}

%
%
%
%
%

\numberofauthors{5} 
\author{
%
%
\alignauthor
Cristian Reyes\\
       \affaddr{Insight Center for Data Analytics}\\
       \affaddr{Dublin, Ireland}\\
       \affaddr{cristian.reyes@estudiant.upc.edu}
\alignauthor
Eva Mohedano\\
       \affaddr{Insight Center for Data Analytics}\\
       \affaddr{Dublin, Ireland}\\
       \affaddr{eva.mohedano@insight-centre.org}
\alignauthor 
Kevin McGuinness\\
       \affaddr{Insight Center for Data Analytics}\\
       \affaddr{Dublin, Ireland}\\
       \affaddr{kevin.mcguinness@dcu.ie}
\and  
\alignauthor 
Noel E. O'Connor\\
       \affaddr{Insight Center for Data Analytics}\\
       \affaddr{Dublin, Ireland}\\
       \affaddr{noel.oconnor@dcu.ie}
\alignauthor Xavier Giro-i-Nieto\\
       \affaddr{Universitat Politecnica de Catalunya}\\
       \affaddr{Barcelona, Catalonia/Spain}\\
       \affaddr{xavier.giro@upc.edu}
}


\maketitle

\input{0_abstract}



\input{1_0_introduction}
\input{2_0_literature}

\input{3_0_methodology}
\input{4_0_experiments}

\input{5_0_conclusions}

\input{acknowledgements}

\bibliographystyle{plain}
\bibliography{sigproc}

%
%

\ccsdesc[100]{Human-centered computing~Ubiquitous and mobile computing systems and tools}

\end{document}

%% file: 0_abstract.tex
\begin{abstract}
This work presents a retrieval pipeline and evaluation scheme for the problem of finding the last appearance of personal objects in a large dataset of images captured from a wearable camera. Each personal object is modelled by a small set of images that define a query for a visual search engine.The retrieved results are reranked considering the temporal timestamps of the images to increase the relevance of the later detections. Finally, a temporal interleaving of the results is introduced for robustness against false detections. The Mean Reciprocal Rank is proposed as a metric to evaluate this problem. This application could help into developing personal assistants capable of helping users when they do not remember where they left their personal belongings.
\end{abstract}

%% file: 1_0_introduction.tex
\section{Introduction}

The interest of users in having their lives digitally recorded has grown in recent years thanks to the advances on wearable sensors, egocentric cameras being among the most informative ones. 
Since wearable cameras are mounted on the user, they are ideal for gathering visual information from everyday interactions.


People interact with their personal belongings several times over the course of the day and, in many cases, they are unintentionally lost because users forget the last location where they were handled. 
Wearable cameras can help users to retrieve candidate locations where the object could be, as the forward egocentric view of these cameras often captures the manipulations with personal belongings as well enough pixels from the background to quickly identify the location. 
Adopting a computer vision approach for finding lost objects is more versatile than a sensor-based one, where some sort of tracking device must be explicitly attached to each object. Cameras do not need any additional device nor an explicit registration of the object. 

There has already been an extensive work on personal object detection in vision, especially when considering video surveillance cameras from CCTV \cite{nascimento2006performance}. 
Wearable cameras offer two main advantages with respect to solutions based on CCTV footage. Firstly, they move together with the user, so their range is not restricted to a specific location. Secondly, a single device can be highly resilient to the object occlusions, as the camera normally takes the same point of view of the user, while a CCTV system will require multiple cameras to cover all points of views.
However, and similarly to CCTV-based systems, these capturing devices typically generate very large volumes of images daily, so finding the relevant images to solve the problem is not a simple task.
An automatic, efficient and scalable retrieval system is necessary to address these situations.

In this work, we assess the potential of egocentric vision to help the user answer the question \textit{Where did I leave my ...?}. 
We address it as a \textit{time-sensitive image retrieval} problem. 
We explore the design of a retrieval system for this purpose, focusing on the visual as well as on the temporal information.
This application could help in developing personal assistants capable of helping users when they do not remember where they left their personal belongings.

This paper focuses in the problem of personal object retrieval from egocentric images with the following contributions:

\begin{itemize}

\item A reranking strategy based on temporal interleaving of those candidate images to contain the query object.


\item A comparison between center bias and saliency maps for the spatial weighting of visual features in the target database. 

\item A proposal of the Mean Reciprocal Rank (MRR) as evaluation metric. 


Explore the usefulness of visual saliency maps  to improve the performance of the visual search engine.
\end{itemize}

The paper is structured as follows. Section \ref{sec:RelatedWork} provides an overview of classic applications for lifelogging data together with previous works that have addressed the problem of finding lost objects with CCTV footage. Section \ref{sec:methodology} presents the proposed solution based on egocentric images, which is based on an off-the-shelf visual search whose retrieved occurrences are reranked to introduce temporal diversity. Section \ref{sec:experiments} evaluates the proposed solution with the Mean Reciprocal Rank (MRR), which is proposed as the reference metric to evaluate personal object retrieval systems. Finally, Section \ref{sec:conclusions} presents the conclusions and draws future research directions.

%% file: 2_0_literature.tex
\section{Related Work}
\label{sec:RelatedWork}

The use of wearable cameras to create persistent personal memories has been tightly associated to the concept of \emph{lifelogging}~\cite{gurrin2014lifelogging}.
Early applications of these devices have focused in healthcare~\cite{doherty2013wearable,gurrin2013smartphone}, with works exploring their applications for patients with mild dementia~\cite{doherty2012experiences}.
More recent works have also explored their applications in affective computing~\cite{hernandez2014senseglass}, social interactions~\cite{Yonetani_2016_CVPR}, and activity recognition~\cite{Ma_2016_CVPR}.





The detection of objects in visual lifelogs has been explored with different applications.
A very popular one is dietary analysis based on the food captured by the camera~\cite{o2013using, bolanos2013active}.
Human-object interactions have been recognized by combining object recognition, motion estimation, and semantic information~\cite{ryoo}, and also by using the hand-object interaction~\cite{gupta, Zhou_2016_CVPR}.
Object recognition is not only useful for object-based detections but also for event identification using the object categories that appear in an image~\cite{lifeifei} or activity recognition based on the object's frequency of use~\cite{wu}. The identification of the active object in the scene was explored in \cite{buso2015geometrical} with the help of visual saliency models.



While object detection and recognition techniques are relevant for this work, in our application we address a retrieval problem, where a ranked list of images from database is shown to the user to help him/her locate their lost object. Previous works from lifelogging have defined retrieval problems for events \cite{doherty2008combining}, audio pieces \cite{shah2012lifelogging}, as well as summarization \cite{chandrasekhar2014efficient} and novelty-detection \cite{aghazadeh2011novelty}.


The problem of retrieval is not only solved with a search based on a similarity metric, as it often also requires the introduction of the notion of diversity. 
The text-based seminal work of  Carbonell \& Goldstain \cite{carbonell1998use} recognizes that  pure relevance ranking  is  not sufficient, so the authors proposed a reranking method that combines independent measurements of relevance and diversity into a single metric to maximize. In image retrieval field, diversification has also shown to increase user satisfaction in ranked results \cite{song2006diversifying}.
Diversity in social image retrieval was one of the focus of the MediaEval benchmarks \cite{ionescu2016result} benchmarks and attracted the interest of many groups working in this field.  
In our problem of personal object retrieval, diversity is especially important because we are mainly interested in the location of the object, more than the object itself. That is, our system should not provide the higher qualities appearances of the object in the database but give hints to the user about the location where the object was last seen.

%% file: 3_0_methodology.tex
\section{Methodology}
\label{sec:methodology}

Our goal is to rank the images captured by a wearable camera during a day based on their likelihood to depict the location of a personal object. 
In our problem we have defined the following sets of images as inputs to the system:

\begin{itemize}
\item The \textbf{query set} $Q$: For each object to be retrieved, a set of exemplar images containing the object is necessary to define the query for the system.

\item The \textbf{target set} $I$: Dataset of images captured by the wearable camera. For each day, this set contains 2,000 images captured throughout the day. 
\end{itemize}

\begin{figure}[!t]
\centering
\includegraphics[width=\columnwidth]{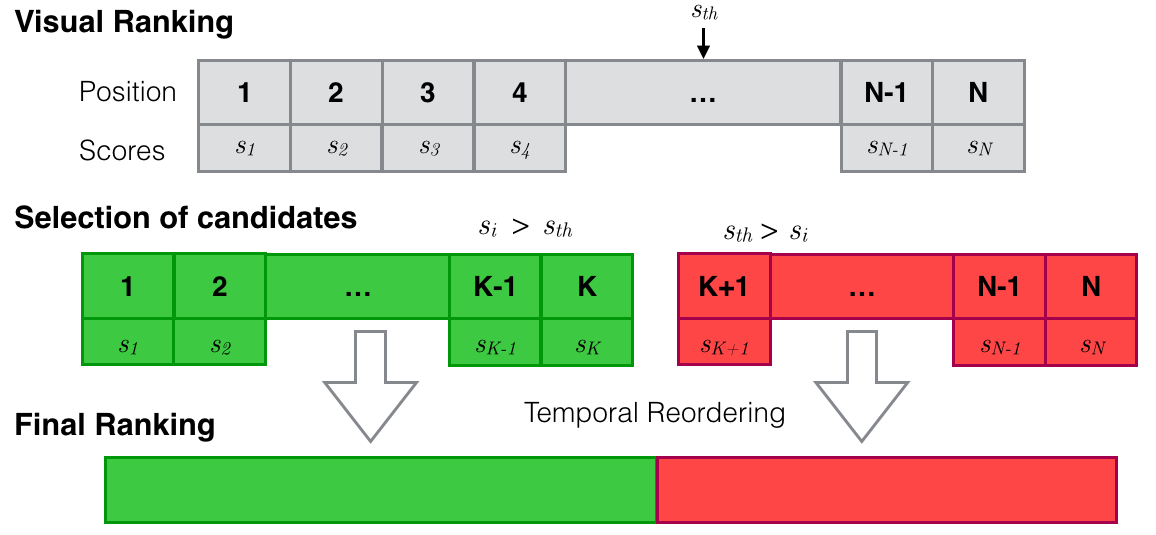}
\caption{Global architecture of the pipeline based system.}
\label{fig:systemSchematic}
\end{figure}

Figure~\ref{fig:systemSchematic} depicts the system architecture. It can be divided in two main blocks: a visually sensitive ranking one followed by a temporally sensitive one. 
The visual block is based on using a pretrained deep convolutional neural network to generate a local representations of an image, and encoding these using bag of words aggregation~\cite{eva2016bags}. At query time, each image in the target set is assigned a score representing its visual similarity with the query object.\\
The temporal block is composed by a first step that selects candidate images and a second one which enhances the ranking using  temporal information.
We included configuration flags for each stage to determine the most appropriate set up. 


\input{3_1_baseline}

\input{3_2_visual_ranking}

\input{3_3_candidate}

\input{3_4_temporal_reranking}

%% file: 3_1_baseline.tex
\subsection{Baseline}
As far as we are aware, there is no previous work on finding lost objects in egocentric images. 
We decided to define as a baseline the simplest approach for the resolution of the problem: a simple temporal sorting of the images based on their time stamp, with the first image in the ranking being the last one taken by the camera.
This mimics the case of a user sequentially browsing through the full sequence of images  for a day in reverse order, the obvious course of action for someone seeking a lost personal item.


%% file: 3_2_visual_ranking.tex
\subsection{Ranking by visual similarity}
\label{sec:visualranking}

The goal of this stage is to create a ranking $R_v$ of the $I$ set for a given $Q$ set. 
The ranking is based only on the visual information of the images.
We explored several configurations and variations of a Convolutional Neural Network Bag-of-Words (BoW) similarity model proposed in \cite{eva2016bags}.
This model is based on the off-the-shelf features learned with the VGG16 network \cite{simonyan2014very} trained on ImageNet \cite{deng2009imagenet}, so no feature learning nor fine-tunning was applied.

The BoW is popular in the image retrieval community and is the basis for some of the best performing techniques of the TRECVID Instance Search Task 2015, where an object instance must be found in a large dataset of videos. Their easy indexing and implementation as an inverted file or multiplication of sparse matrixes makes it a very common approach for content-based object retrieval systems.

\subsubsection{Encoding the query images}
\label{subsec:encodingImageQuery}
\begin{figure}[ht]
\centering
\includegraphics[width=0.45\textwidth]{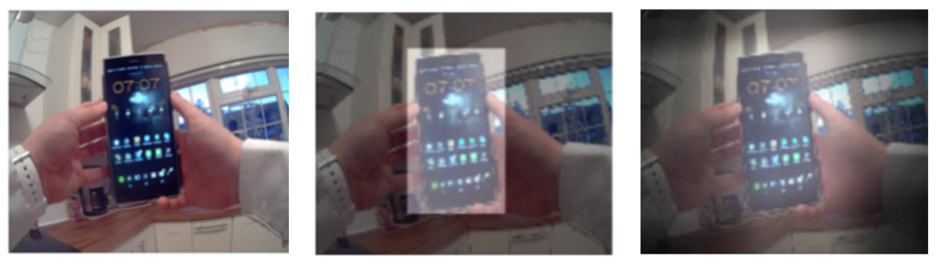}

\caption[Encoding the query images]{Examples of the different masking strategies applied in a query image. Left: Full image, center: Hard Bounding Box, Right: Soft Bounding Box.}
\label{fig:queryEncoding}
\end{figure}

A function $f:  Q  \longmapsto f(Q) \in \mathbb{R}^n$ aims at building a single \textbf{query vector} $\vec{q}$ by gathering the information of all images in $Q = \left\lbrace q_1, q_2,...,q_{|Q|} \right\rbrace$ to obtain $\vec{q} = f(Q)$. 
Three different approaches have been explored to define $f$, illustrated in Figure~\ref{fig:queryEncoding}:

\textbf{a) Full Image (FI)}: The $\vec{q}$ vector is constructed by averaging the frequencies of the visual words of all the local CNN features from the query images.

\textbf{b) Hard Bounding Box (HBB)}: The $\vec{q}$ vector is constructed by averaging frequencies of the visual words that fall inside a query bounding box that surrounds the object. This approach considers only the visual words that describe the object.

\textbf{c) Soft Bounding Box (SBB)}: The $\vec{q}$ vector is constructed by averaging frequencies of the visual words of the whole image, but weighting them depending on their distance to the bounding box. This allows introducing context in addition to the object. Weights are computed as the inverse of the distance to the closest side of the bounding box and are $L_2$-normalized.

\begin{figure}[ht]
\centering
\includegraphics[width=0.45\textwidth]{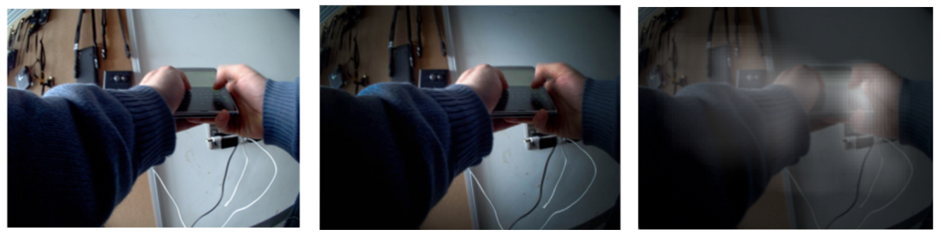}

\caption[Encoding the target images]{Examples of the different masking strategies applied in a target image. Left: Full image, center: Center Bias, Right: Saliency Mask.}
\label{fig:targetEncoding}
\end{figure}

\subsubsection{Encoding the target images}
\label{subsec:encodingImageQuery}
A similar procedure is applied to the set of target images $I$, the daily images in our problem. A function $g:I \rightarrow \mathbb{R}^n$ is defined to build a feature vector $\vec{i}_j = g(i_j)$ for each image $i_j \in I$. Three different definitions of the $g$ function have been studied (Figure~\ref{fig:targetEncoding}):

\textbf{a) Full Image (FI)}: The $\vec{i_j}$ vector is built using the visual words of all the local CNN features from the $i_j$ image.

\textbf{b) Center Bias (CB)}: The $\vec{i_j}$ vector is built using the visual words of all the local CNN features from the $i_j$ image but it inversely weightens the features with the distance to the center of the image.
This approach is inspired by previous works in the field of salient object detection \cite{liu2011learning}.

\textbf{c) Saliency Mask (SM)}: The $\vec{i_j}$ vector is built using the local CNN features of the whole image, but this time weighting their frequencies using a saliency map generated using a computational model of visual saliency.
Using saliency maps for object detection  and recognition has been previously proposed in \cite{hu2014flower, giouvanakis2014saliency, buso2015geometrical}.

Following~\cite{eva2016bags}, an assignment map is extracted from the \textit{conv-5\_1} layer of the \textit{VGG-16} pre-trained convolutional neural network, giving a $32\times 42$ assignment map.

Saliency maps are calculated for each image using the pre-trained \textit{SalNet}~\cite{SalNet} CNN. This network produces maps that represent the probability of visual attention on an image, defined as the eye gaze fixation points. We downsample the saliency maps to match the size of the assignment maps by average pooling over local blocks. 
After downsampling a vector of weights $w = (w_1, ..., w_{32 \times 42})$ is constructed and $L_2$-normalized. 





Finally, each $\vec{i}_j$ feature vector is compared with the $\vec{q}$ query feature vector using cosine similarity~\footnote{
$\textnormal{cosine similarity}$($a$,$b$) = cos($\widehat{a,b}$) Note that it is always between 0 and 1 as vectors have non-negative components.
} 
between $\vec{i}$ and $\vec{q}$ and obtain the $\nu$ score.
Then the visual ranking $R_v$ is produced by ordering the images in $I$ according to their $\nu$ score.

%% file: 3_3_candidate.tex
\subsection{Detection of candidate moments}
\label{sec:candidate}

The visual ranking $R_v$ provides an ordered list of the images based on their likelihood to contain the object. 
Notice that in our problem this information might not always be useful. 
The last appearance of the object does not need to be the most similar to the query in visual terms. 
Taking this into account we introduced a post-processing to the visual search ranking.

The first step in this post-processing is determining which of the images in the ranked list should be considered as likely to contain the query object. 
This is achieved by thresholding the list and considering as candidate images ($C$) those with scores higher than the threshold, and discarded images ($D$) those with scores below it.
Two different thresholding techniques were considered in order to create the $C$ and $D = I \setminus C$ sets. 

\textbf{a) Threshold on Visual Similarity Scores} (TVSS): This technique consists in building the set of the candidate images as $C = \left\lbrace p \in I : \nu_p > \nu_{th} \right\rbrace$, where $\nu_{th}$ is a learned threshold. It is an absolute threshold that the visual scores have to overcome to be considered as candidates.

\textbf{b) Nearest Neighbor Distance Ratio} (NDRR): This strategy is inspired by Lowe~\cite{loewe2004}.
Let $\nu_1$ and $\nu_2$ be the two best scores in the ranked list, then the candidates set is defined as \[
C = \left\lbrace i \in I : \frac{\nu_i}{\nu_1} > \rho_{th} \frac{\nu_2}{\nu_1} \right\rbrace.
\]
In this case, it is an adaptive technique that sets the threshold depending on the ratio of the scores of two best visually ranked images.

Both techniques require to set either $\nu_{th}$ or $\rho_{th}$. These values cannot be chosen arbitrarily, so they were learned from a training process, described in section \ref{sec:training}. 

%% file: 3_4_temporal_reranking.tex
\subsection{Temporally aware reranking}
\label{sec:temporalreranking}

Once candidate images have been selected based on their visual features, the next, and last, step considers the temporal information. A temporal-aware reranking introduces the concept that the lost object may not be in the location with the best visual match with the query, but in the last location where it was seen. 

Two rankings $R_C$ and $R_D$ are built by reranking the elements in $C$ and $D$, respectively, based on their time stamps. The final ranking $R_t$ is built as the concatenation of $R_t = [R_C,R_D]$  (which considers the best candidate to be at the beginning of the sequence). Thus, $R_t$ always contains all the images in $I$ and we ensure that any relevant image will appear somewhere in the ranking, even after the thresholded cases.
We propose two strategies to exploit the time stamps of the images:

\textbf{a) Decreasing Time-Stamp Sorting}: This is the most simple approach we can consider at this point. Just a simple reordering of the $C$ and $D$ sets to build the $R_C$ and $R_D$ rankings from the latest to the earliest time-stamp. This configuration will be applied in all experiments, unless otherwise stated.

\textbf{b) Interleaving}: This other approach introduces the concept of diversity. We realized that the rankings tend to present consecutive images of the same moment when using the straightforward sorting. This is expected behavior due to the high visual redundancy of neighboring images in an egocentric sequence.
As the final goal of this work is determining the location of the object, showing similar and consecutive images to the user is uninformative. By introducing a diversity step, we force the system to generate a rank list of diverse images, which may increase the chances of determining the object location by looking at the minimum of elements in the ranked list.

\begin{figure}[t]
\centering
\includegraphics[width=0.4\textwidth]{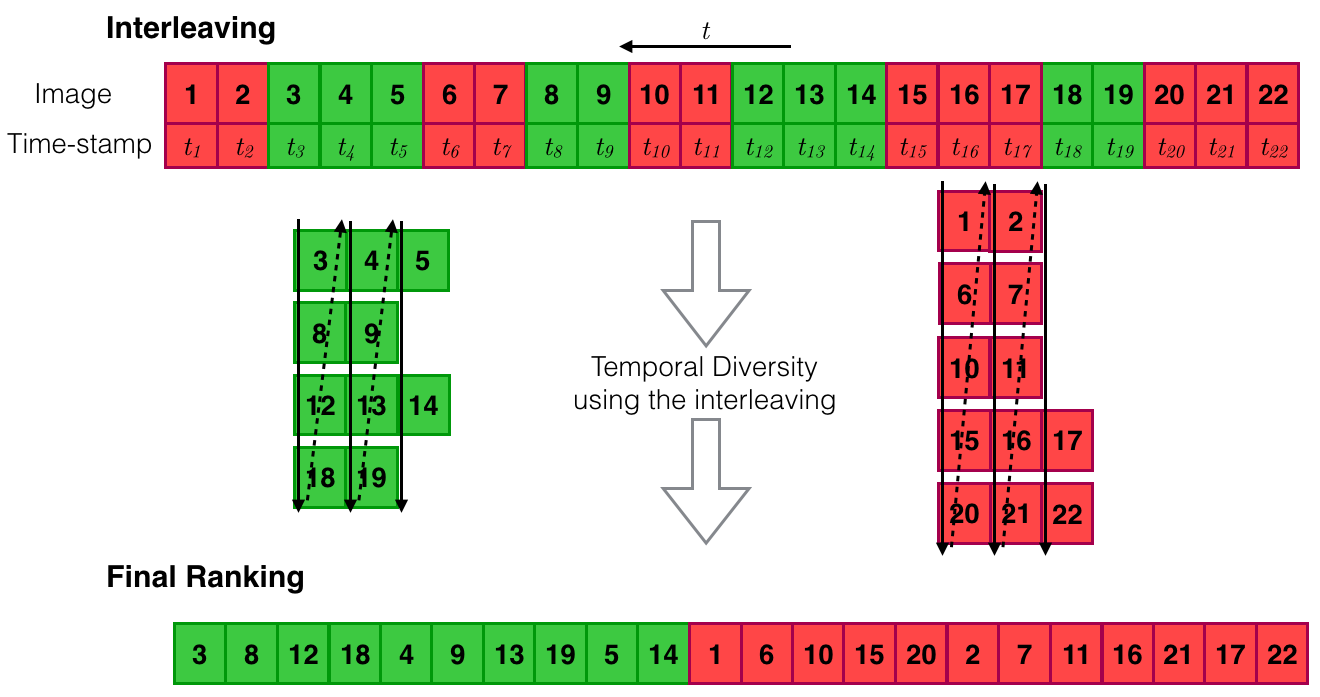}
\caption{Scheme of the interleaving strategy used.}
\label{fig:diversity}
\end{figure}

Our diversity-based technique has its basis in the interleaving of samples. 
In digital communication, interleaving is the reordering of data that is to be transmitted so that consecutive samples are distributed over a larger sequence of data in order to reduce the effect of burst errors. 
Adapting it to our domain, we interleave images from different scenes to put a representative of each scene early in the ranking. 
Thus, if the first candidate is not relevant, we avoid the second to be from the same scene and, therefore, it is more likely to be relevant. Figure~\ref{fig:diversity} depicts the temporal diversity strategy. The algorithm proceeds as follows:
\begin{enumerate}
\item Make a list with all the images in $I$ sorting by their time-stamp in decreasing order. That is, the later image the first. For each image $i \in I$ it must be known whether it belongs to $C$ or $D$. Such as, $$O=\left\lbrace i_{n-1}^C, ..., i_{m}^C,  i_{m-1}^D, ..., i_{l}^D, i_{l-1}^C, ..., i_{k}^C,  i_{k-1}^D, ..., i_{1}^{D} \right\rbrace$$

\item Split into sub-lists using the transitions $C \rightarrow D$ or $D \rightarrow C$ as a boundary.

\item Build a new list $R_C$ by adding the first image of each sub-list containing elements in $C$ maintaining time-stamp in decreasing order. Then, the second image of each sub-list and so on. Thus, $R_C=\left\lbrace i_{n-1}^C, i_{l-1}^C, i_{n-2}^C, i_{l-2}^C, ... \right\rbrace$. Build $R_D$ analogously.

\item Concatenate $R_C$ and $R_D$ to obtain the final ranking $R_t = [R_C,R_D]$.
\end{enumerate}


\input{graph_training}

%% file: graph_training.tex
\FloatBarrier
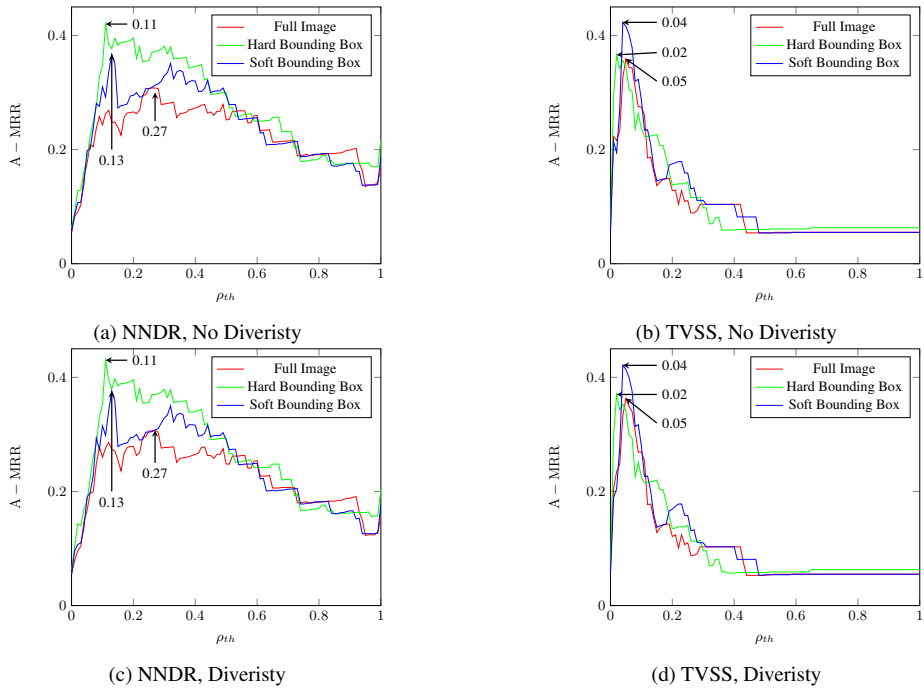
\begin{figure*}[h]
\centering
\begin{subfigure}[b]{0.4\textwidth}
  \centering
  \begin{tikzpicture}[scale=0.6]
    \begin{axis}[
        name=plot1,
        title={},
        xlabel={$\rho_{th}$},
        ylabel={$\mathrm{A-MRR}$},
        xmin=0, xmax=1,
        ymin=0, ymax=0.45,
        xtick={0,0.2,0.40,0.60,0.80,1.00},
        ytick={0,0.20,0.40,0.60,0.80,1.00},
        legend pos=north east,
        ymajorgrids=false,
        grid style=dashed,
    ]

    \addplot
        [color=red]
        coordinates {
        (0.0,0.055) 
        (0.01,0.083) 
        (0.02,0.091) 
        (0.03,0.103) 
        (0.04,0.151) 
        (0.05,0.176) 
        (0.06,0.209) 
        (0.07,0.205) 
        (0.08,0.237) 
        (0.09,0.258) 
        (0.1,0.243) 
        (0.11,0.263) 
        (0.12,0.269) 
        (0.13,0.248) 
        (0.14,0.248) 
        (0.15,0.239) 
        (0.16,0.225) 
        (0.17,0.257) 
        (0.18,0.264) 
        (0.19,0.265) 
        (0.2,0.266) 
        (0.21,0.263) 
        (0.22,0.269) 
        (0.23,0.295) 
        (0.24,0.296) 
        (0.25,0.307) 
        (0.26,0.308) 
        (0.27,0.308) 
        (0.28,0.307) 
        (0.29,0.279) 
        (0.3,0.28) 
        (0.31,0.281) 
        (0.32,0.282) 
        (0.33,0.283) 
        (0.34,0.256) 
        (0.35,0.263) 
        (0.36,0.264) 
        (0.37,0.265) 
        (0.38,0.27) 
        (0.39,0.271) 
        (0.4,0.273) 
        (0.41,0.275) 
        (0.42,0.278) 
        (0.43,0.279) 
        (0.44,0.264) 
        (0.45,0.266) 
        (0.46,0.26) 
        (0.47,0.267) 
        (0.48,0.267) 
        (0.49,0.279) 
        (0.5,0.252) 
        (0.51,0.253) 
        (0.52,0.267) 
        (0.53,0.267) 
        (0.54,0.268) 
        (0.55,0.268) 
        (0.56,0.268) 
        (0.57,0.247) 
        (0.58,0.247) 
        (0.59,0.259) 
        (0.6,0.26) 
        (0.61,0.233) 
        (0.62,0.234) 
        (0.63,0.234) 
        (0.64,0.235) 
        (0.65,0.213) 
        (0.66,0.213) 
        (0.67,0.214) 
        (0.68,0.214) 
        (0.69,0.215) 
        (0.7,0.215) 
        (0.71,0.216) 
        (0.72,0.216) 
        (0.73,0.189) 
        (0.74,0.19) 
        (0.75,0.19) 
        (0.76,0.192) 
        (0.77,0.189) 
        (0.78,0.19) 
        (0.79,0.191) 
        (0.8,0.191) 
        (0.81,0.192) 
        (0.82,0.192) 
        (0.83,0.193) 
        (0.84,0.193) 
        (0.85,0.193) 
        (0.86,0.194) 
        (0.87,0.194) 
        (0.88,0.196) 
        (0.89,0.199) 
        (0.9,0.199) 
        (0.91,0.201) 
        (0.92,0.202) 
        (0.93,0.176) 
        (0.94,0.163) 
        (0.95,0.135) 
        (0.96,0.138) 
        (0.97,0.138) 
        (0.98,0.138) 
        (0.99,0.143) 
        (1.0,0.167) 
        };
        \addlegendentry{Full Image}
        \draw [stealth-,thick] (axis cs:0.27,0.3) -- (axis cs:0.27,0.25) node[below]{0.27} ;
        
        \addplot
        [color=green]
        coordinates {
        (0.0,0.063) 
        (0.01,0.09) 
        (0.02,0.128) 
        (0.03,0.129) 
        (0.04,0.163) 
        (0.05,0.184) 
        (0.06,0.223) 
        (0.07,0.248) 
        (0.08,0.275) 
        (0.09,0.328) 
        (0.1,0.347) 
        (0.11,0.42) 
        (0.12,0.385) 
        (0.13,0.377) 
        (0.14,0.395) 
        (0.15,0.383) 
        (0.16,0.388) 
        (0.17,0.388) 
        (0.18,0.39) 
        (0.19,0.391) 
        (0.2,0.396) 
        (0.21,0.36) 
        (0.22,0.38) 
        (0.23,0.354) 
        (0.24,0.355) 
        (0.25,0.356) 
        (0.26,0.37) 
        (0.27,0.372) 
        (0.28,0.373) 
        (0.29,0.376) 
        (0.3,0.381) 
        (0.31,0.368) 
        (0.32,0.38) 
        (0.33,0.359) 
        (0.34,0.36) 
        (0.35,0.36) 
        (0.36,0.361) 
        (0.37,0.362) 
        (0.38,0.343) 
        (0.39,0.344) 
        (0.4,0.346) 
        (0.41,0.347) 
        (0.42,0.349) 
        (0.43,0.322) 
        (0.44,0.324) 
        (0.45,0.297) 
        (0.46,0.297) 
        (0.47,0.298) 
        (0.48,0.299) 
        (0.49,0.3) 
        (0.5,0.3) 
        (0.51,0.273) 
        (0.52,0.273) 
        (0.53,0.258) 
        (0.54,0.259) 
        (0.55,0.259) 
        (0.56,0.262) 
        (0.57,0.262) 
        (0.58,0.263) 
        (0.59,0.25) 
        (0.6,0.25) 
        (0.61,0.25) 
        (0.62,0.25) 
        (0.63,0.251) 
        (0.64,0.251) 
        (0.65,0.256) 
        (0.66,0.257) 
        (0.67,0.257) 
        (0.68,0.231) 
        (0.69,0.232) 
        (0.7,0.232) 
        (0.71,0.232) 
        (0.72,0.204) 
        (0.73,0.206) 
        (0.74,0.18) 
        (0.75,0.179) 
        (0.76,0.18) 
        (0.77,0.181) 
        (0.78,0.182) 
        (0.79,0.182) 
        (0.8,0.184) 
        (0.81,0.189) 
        (0.82,0.19) 
        (0.83,0.179) 
        (0.84,0.181) 
        (0.85,0.174) 
        (0.86,0.174) 
        (0.87,0.175) 
        (0.88,0.175) 
        (0.89,0.175) 
        (0.9,0.175) 
        (0.91,0.176) 
        (0.92,0.176) 
        (0.93,0.176) 
        (0.94,0.176) 
        (0.95,0.176) 
        (0.96,0.177) 
        (0.97,0.17) 
        (0.98,0.17) 
        (0.99,0.17) 
        (1.0,0.217) 
        };
        \addlegendentry{Hard Bounding Box}
        \draw [stealth-,thick] (axis cs:0.11,0.420) -- (axis cs:0.18,0.420) node[right]{0.11} ;

    \addplot
        [color=blue]
        coordinates {
          (0.0,0.055) 
          (0.01,0.086) 
          (0.02,0.107) 
          (0.03,0.108) 
          (0.04,0.14) 
          (0.05,0.198) 
          (0.06,0.201) 
          (0.07,0.237) 
          (0.08,0.281) 
          (0.09,0.276) 
          (0.1,0.31) 
          (0.11,0.292) 
          (0.12,0.335) 
          (0.13,0.367) 
          (0.14,0.351) 
          (0.15,0.273) 
          (0.16,0.277) 
          (0.17,0.278) 
          (0.18,0.28) 
          (0.19,0.294) 
          (0.2,0.295) 
          (0.21,0.3) 
          (0.22,0.292) 
          (0.23,0.293) 
          (0.24,0.307) 
          (0.25,0.307) 
          (0.26,0.309) 
          (0.27,0.313) 
          (0.28,0.315) 
          (0.29,0.32) 
          (0.3,0.33) 
          (0.31,0.335) 
          (0.32,0.351) 
          (0.33,0.324) 
          (0.34,0.336) 
          (0.35,0.339) 
          (0.36,0.336) 
          (0.37,0.314) 
          (0.38,0.319) 
          (0.39,0.321) 
          (0.4,0.32) 
          (0.41,0.302) 
          (0.42,0.302) 
          (0.43,0.307) 
          (0.44,0.321) 
          (0.45,0.3) 
          (0.46,0.302) 
          (0.47,0.303) 
          (0.48,0.303) 
          (0.49,0.308) 
          (0.5,0.294) 
          (0.51,0.279) 
          (0.52,0.279) 
          (0.53,0.279) 
          (0.54,0.253) 
          (0.55,0.253) 
          (0.56,0.254) 
          (0.57,0.254) 
          (0.58,0.255) 
          (0.59,0.255) 
          (0.6,0.256) 
          (0.61,0.229) 
          (0.62,0.23) 
          (0.63,0.208) 
          (0.64,0.209) 
          (0.65,0.209) 
          (0.66,0.209) 
          (0.67,0.21) 
          (0.68,0.21) 
          (0.69,0.211) 
          (0.7,0.212) 
          (0.71,0.213) 
          (0.72,0.214) 
          (0.73,0.214) 
          (0.74,0.187) 
          (0.75,0.188) 
          (0.76,0.188) 
          (0.77,0.19) 
          (0.78,0.191) 
          (0.79,0.192) 
          (0.8,0.192) 
          (0.81,0.193) 
          (0.82,0.193) 
          (0.83,0.193) 
          (0.84,0.172) 
          (0.85,0.171) 
          (0.86,0.171) 
          (0.87,0.173) 
          (0.88,0.173) 
          (0.89,0.176) 
          (0.9,0.176) 
          (0.91,0.176) 
          (0.92,0.162) 
          (0.93,0.162) 
          (0.94,0.137) 
          (0.95,0.138) 
          (0.96,0.138) 
          (0.97,0.138) 
          (0.98,0.138) 
          (0.99,0.139) 
          (1.0,0.189) 
        };
        \addlegendentry{Soft Bounding Box}
        \draw [stealth-,thick] (axis cs:0.13,0.367) -- (axis cs:0.13,0.2) node[below]{0.13} ;

    \end{axis}
    ;
  \end{tikzpicture}
  \caption[Ex2]%
  {{\small NNDR, No Diveristy}}    
\end{subfigure}
\begin{subfigure}[b]{0.4\textwidth}
  \centering
\begin{tikzpicture}[scale=0.6]
    \begin{axis}[
        name=plot1,
        title={},
        xlabel={$\rho_{th}$},
        ylabel={$\mathrm{A-MRR}$},
        xmin=0, xmax=1,
        ymin=0, ymax=0.45,
        xtick={0,0.2,0.40,0.60,0.80,1.00},
        ytick={0,0.20,0.40,0.60,0.80,1.00},
        legend pos=north east,
        ymajorgrids=false,
        grid style=dashed,
    ]

    \addplot
        [color=red]
        coordinates {
          (0.0,0.055) 
          (0.01,0.223) 
          (0.02,0.216) 
          (0.03,0.233) 
          (0.04,0.326) 
          (0.05,0.359) 
          (0.06,0.344) 
          (0.07,0.343) 
          (0.08,0.308) 
          (0.09,0.275) 
          (0.1,0.275) 
          (0.11,0.26) 
          (0.12,0.185) 
          (0.13,0.187) 
          (0.14,0.161) 
          (0.15,0.137) 
          (0.16,0.143) 
          (0.17,0.148) 
          (0.18,0.149) 
          (0.19,0.15) 
          (0.2,0.128) 
          (0.21,0.129) 
          (0.22,0.104) 
          (0.23,0.128) 
          (0.24,0.109) 
          (0.25,0.111) 
          (0.26,0.089) 
          (0.27,0.089) 
          (0.28,0.093) 
          (0.29,0.104) 
          (0.3,0.104) 
          (0.31,0.104) 
          (0.32,0.104) 
          (0.33,0.104) 
          (0.34,0.104) 
          (0.35,0.104) 
          (0.36,0.104) 
          (0.37,0.104) 
          (0.38,0.104) 
          (0.39,0.104) 
          (0.4,0.104) 
          (0.41,0.104) 
          (0.42,0.104) 
          (0.43,0.076) 
          (0.44,0.054) 
          (0.45,0.054) 
          (0.46,0.054) 
          (0.47,0.054) 
          (0.48,0.054) 
          (0.49,0.054) 
          (0.5,0.054) 
          (0.51,0.054) 
          (0.52,0.054) 
          (0.53,0.055) 
          (0.54,0.055) 
          (0.55,0.055) 
          (0.56,0.055) 
          (0.57,0.055) 
          (0.58,0.055) 
          (0.59,0.055) 
          (0.6,0.055) 
          (0.61,0.055) 
          (0.62,0.055) 
          (0.63,0.055) 
          (0.64,0.055) 
          (0.65,0.055) 
          (0.66,0.055) 
          (0.67,0.055) 
          (0.68,0.055) 
          (0.69,0.055) 
          (0.7,0.055) 
          (0.71,0.055) 
          (0.72,0.055) 
          (0.73,0.055) 
          (0.74,0.055) 
          (0.75,0.055) 
          (0.76,0.055) 
          (0.77,0.055) 
          (0.78,0.055) 
          (0.79,0.055) 
          (0.8,0.055) 
          (0.81,0.055) 
          (0.82,0.055) 
          (0.83,0.055) 
          (0.84,0.055) 
          (0.85,0.055) 
          (0.86,0.055) 
          (0.87,0.055) 
          (0.88,0.055) 
          (0.89,0.055) 
          (0.9,0.055) 
          (0.91,0.055) 
          (0.92,0.055) 
          (0.93,0.055) 
          (0.94,0.055) 
          (0.95,0.055) 
          (0.96,0.055) 
          (0.97,0.055) 
          (0.98,0.055) 
          (0.99,0.055) 
          (1.0,0.055) 
        };
        \addlegendentry{Full Image}
        \draw [stealth-,thick] (axis cs:0.05,0.359) -- (axis cs:0.15,0.32) node[right]{0.05} ;
        
        \addplot
        [color=green]
        coordinates {
        (0.0,0.063) 
        (0.01,0.294) 
        (0.02,0.366) 
        (0.03,0.342) 
        (0.04,0.356) 
        (0.05,0.346) 
        (0.06,0.306) 
        (0.07,0.302) 
        (0.08,0.237) 
        (0.09,0.263) 
        (0.1,0.237) 
        (0.11,0.223) 
        (0.12,0.223) 
        (0.13,0.224) 
        (0.14,0.225) 
        (0.15,0.226) 
        (0.16,0.207) 
        (0.17,0.207) 
        (0.18,0.186) 
        (0.19,0.16) 
        (0.2,0.138) 
        (0.21,0.139) 
        (0.22,0.14) 
        (0.23,0.14) 
        (0.24,0.14) 
        (0.25,0.143) 
        (0.26,0.116) 
        (0.27,0.116) 
        (0.28,0.116) 
        (0.29,0.116) 
        (0.3,0.098) 
        (0.31,0.098) 
        (0.32,0.07) 
        (0.33,0.07) 
        (0.34,0.081) 
        (0.35,0.081) 
        (0.36,0.059) 
        (0.37,0.059) 
        (0.38,0.059) 
        (0.39,0.059) 
        (0.4,0.059) 
        (0.41,0.06) 
        (0.42,0.06) 
        (0.43,0.06) 
        (0.44,0.06) 
        (0.45,0.06) 
        (0.46,0.06) 
        (0.47,0.06) 
        (0.48,0.06) 
        (0.49,0.06) 
        (0.5,0.06) 
        (0.51,0.06) 
        (0.52,0.061) 
        (0.53,0.061) 
        (0.54,0.061) 
        (0.55,0.061) 
        (0.56,0.061) 
        (0.57,0.061) 
        (0.58,0.061) 
        (0.59,0.061) 
        (0.6,0.061) 
        (0.61,0.061) 
        (0.62,0.061) 
        (0.63,0.061) 
        (0.64,0.061) 
        (0.65,0.063) 
        (0.66,0.063) 
        (0.67,0.063) 
        (0.68,0.063) 
        (0.69,0.063) 
        (0.7,0.063) 
        (0.71,0.063) 
        (0.72,0.063) 
        (0.73,0.063) 
        (0.74,0.063) 
        (0.75,0.063) 
        (0.76,0.063) 
        (0.77,0.063) 
        (0.78,0.063) 
        (0.79,0.063) 
        (0.8,0.063) 
        (0.81,0.063) 
        (0.82,0.063) 
        (0.83,0.063) 
        (0.84,0.063) 
        (0.85,0.063) 
        (0.86,0.063) 
        (0.87,0.063) 
        (0.88,0.063) 
        (0.89,0.063) 
        (0.9,0.063) 
        (0.91,0.063) 
        (0.92,0.063) 
        (0.93,0.063) 
        (0.94,0.063) 
        (0.95,0.063) 
        (0.96,0.063) 
        (0.97,0.063) 
        (0.98,0.063) 
        (0.99,0.063) 
        (1.0,0.063) 
        };
        \addlegendentry{Hard Bounding Box}
        \draw [stealth-,thick] (axis cs:0.02,0.366) -- (axis cs:0.15,0.37) node[right]{0.02} ;

    \addplot
        [color=blue]
        coordinates {
          (0.0,0.055) 
          (0.01,0.214) 
          (0.02,0.195) 
          (0.03,0.271) 
          (0.04,0.423) 
          (0.05,0.416) 
          (0.06,0.402) 
          (0.07,0.379) 
          (0.08,0.32) 
          (0.09,0.323) 
          (0.1,0.287) 
          (0.11,0.234) 
          (0.12,0.235) 
          (0.13,0.209) 
          (0.14,0.168) 
          (0.15,0.145) 
          (0.16,0.147) 
          (0.17,0.148) 
          (0.18,0.149) 
          (0.19,0.173) 
          (0.2,0.175) 
          (0.21,0.177) 
          (0.22,0.179) 
          (0.23,0.179) 
          (0.24,0.159) 
          (0.25,0.159) 
          (0.26,0.131) 
          (0.27,0.133) 
          (0.28,0.111) 
          (0.29,0.111) 
          (0.3,0.111) 
          (0.31,0.104) 
          (0.32,0.104) 
          (0.33,0.104) 
          (0.34,0.104) 
          (0.35,0.104) 
          (0.36,0.104) 
          (0.37,0.104) 
          (0.38,0.104) 
          (0.39,0.104) 
          (0.4,0.104) 
          (0.41,0.082) 
          (0.42,0.082) 
          (0.43,0.082) 
          (0.44,0.082) 
          (0.45,0.082) 
          (0.46,0.082) 
          (0.47,0.082) 
          (0.48,0.054) 
          (0.49,0.054) 
          (0.5,0.054) 
          (0.51,0.054) 
          (0.52,0.054) 
          (0.53,0.054) 
          (0.54,0.054) 
          (0.55,0.054) 
          (0.56,0.054) 
          (0.57,0.054) 
          (0.58,0.054) 
          (0.59,0.055) 
          (0.6,0.055) 
          (0.61,0.055) 
          (0.62,0.055) 
          (0.63,0.055) 
          (0.64,0.055) 
          (0.65,0.055) 
          (0.66,0.055) 
          (0.67,0.055) 
          (0.68,0.055) 
          (0.69,0.055) 
          (0.7,0.055) 
          (0.71,0.055) 
          (0.72,0.055) 
          (0.73,0.055) 
          (0.74,0.055) 
          (0.75,0.055) 
          (0.76,0.055) 
          (0.77,0.055) 
          (0.78,0.055) 
          (0.79,0.055) 
          (0.8,0.055) 
          (0.81,0.055) 
          (0.82,0.055) 
          (0.83,0.055) 
          (0.84,0.055) 
          (0.85,0.055) 
          (0.86,0.055) 
          (0.87,0.055) 
          (0.88,0.055) 
          (0.89,0.055) 
          (0.9,0.055) 
          (0.91,0.055) 
          (0.92,0.055) 
          (0.93,0.055) 
          (0.94,0.055) 
          (0.95,0.055) 
          (0.96,0.055) 
          (0.97,0.055) 
          (0.98,0.055) 
          (0.99,0.055) 
          (1.0,0.055) 
        };
        \addlegendentry{Soft Bounding Box}
        \draw [stealth-,thick] (axis cs:0.04,0.423) -- (axis cs:0.15,0.423) node[right]{0.04} ;

    \end{axis}
    ;
  \end{tikzpicture}
  \caption[Ex2]%
  {{\small TVSS, No Diveristy}}    
\end{subfigure}

\begin{subfigure}[b]{0.4\textwidth}
  \centering
\begin{tikzpicture}[scale=0.6]
    \begin{axis}[
        name=plot1,
        title={},
        xlabel={$\rho_{th}$},
        ylabel={$\mathrm{A-MRR}$},
        xmin=0, xmax=1,
        ymin=0, ymax=0.45,
        xtick={0,0.2,0.40,0.60,0.80,1.00},
        ytick={0,0.20,0.40,0.60,0.80,1.00},
        legend pos=north east,
        ymajorgrids=false,
        grid style=dashed,
    ]

    \addplot
        [color=red]
        coordinates {
		(0.0,0.055) 
(0.01,0.079) 
(0.02,0.096) 
(0.03,0.104) 
(0.04,0.147) 
(0.05,0.166) 
(0.06,0.197) 
(0.07,0.199) 
(0.08,0.248) 
(0.09,0.278) 
(0.1,0.26) 
(0.11,0.276) 
(0.12,0.286) 
(0.13,0.274) 
(0.14,0.271) 
(0.15,0.255) 
(0.16,0.236) 
(0.17,0.267) 
(0.18,0.277) 
(0.19,0.279) 
(0.2,0.279) 
(0.21,0.263) 
(0.22,0.268) 
(0.23,0.295) 
(0.24,0.296) 
(0.25,0.306) 
(0.26,0.306) 
(0.27,0.307) 
(0.28,0.304) 
(0.29,0.276) 
(0.3,0.277) 
(0.31,0.277) 
(0.32,0.278) 
(0.33,0.278) 
(0.34,0.251) 
(0.35,0.258) 
(0.36,0.259) 
(0.37,0.26) 
(0.38,0.261) 
(0.39,0.263) 
(0.4,0.265) 
(0.41,0.267) 
(0.42,0.277) 
(0.43,0.278) 
(0.44,0.263) 
(0.45,0.263) 
(0.46,0.258) 
(0.47,0.265) 
(0.48,0.265) 
(0.49,0.275) 
(0.5,0.248) 
(0.51,0.248) 
(0.52,0.262) 
(0.53,0.262) 
(0.54,0.262) 
(0.55,0.263) 
(0.56,0.263) 
(0.57,0.241) 
(0.58,0.241) 
(0.59,0.253) 
(0.6,0.254) 
(0.61,0.227) 
(0.62,0.227) 
(0.63,0.227) 
(0.64,0.228) 
(0.65,0.206) 
(0.66,0.206) 
(0.67,0.206) 
(0.68,0.207) 
(0.69,0.207) 
(0.7,0.207) 
(0.71,0.207) 
(0.72,0.207) 
(0.73,0.18) 
(0.74,0.18) 
(0.75,0.181) 
(0.76,0.182) 
(0.77,0.179) 
(0.78,0.18) 
(0.79,0.181) 
(0.8,0.181) 
(0.81,0.182) 
(0.82,0.182) 
(0.83,0.183) 
(0.84,0.183) 
(0.85,0.183) 
(0.86,0.184) 
(0.87,0.184) 
(0.88,0.186) 
(0.89,0.188) 
(0.9,0.188) 
(0.91,0.19) 
(0.92,0.191) 
(0.93,0.164) 
(0.94,0.15) 
(0.95,0.123) 
(0.96,0.124) 
(0.97,0.124) 
(0.98,0.124) 
(0.99,0.129) 
(1.0,0.155) 
        };
        \addlegendentry{Full Image}
        \draw [stealth-,thick] (axis cs:0.27,0.307) -- (axis cs:0.27,0.25) node[below]{0.27} ;
        
        \addplot
        [color=green]
        coordinates {
(0.0,0.063) 
(0.01,0.102) 
(0.02,0.143) 
(0.03,0.14) 
(0.04,0.169) 
(0.05,0.183) 
(0.06,0.222) 
(0.07,0.261) 
(0.08,0.282) 
(0.09,0.331) 
(0.1,0.353) 
(0.11,0.43) 
(0.12,0.395) 
(0.13,0.38) 
(0.14,0.397) 
(0.15,0.385) 
(0.16,0.389) 
(0.17,0.389) 
(0.18,0.391) 
(0.19,0.392) 
(0.2,0.396) 
(0.21,0.359) 
(0.22,0.38) 
(0.23,0.355) 
(0.24,0.355) 
(0.25,0.356) 
(0.26,0.37) 
(0.27,0.37) 
(0.28,0.371) 
(0.29,0.374) 
(0.3,0.379) 
(0.31,0.365) 
(0.32,0.378) 
(0.33,0.356) 
(0.34,0.357) 
(0.35,0.357) 
(0.36,0.358) 
(0.37,0.359) 
(0.38,0.339) 
(0.39,0.339) 
(0.4,0.34) 
(0.41,0.342) 
(0.42,0.344) 
(0.43,0.317) 
(0.44,0.319) 
(0.45,0.291) 
(0.46,0.291) 
(0.47,0.292) 
(0.48,0.293) 
(0.49,0.293) 
(0.5,0.293) 
(0.51,0.266) 
(0.52,0.266) 
(0.53,0.251) 
(0.54,0.252) 
(0.55,0.252) 
(0.56,0.254) 
(0.57,0.255) 
(0.58,0.255) 
(0.59,0.242) 
(0.6,0.242) 
(0.61,0.242) 
(0.62,0.242) 
(0.63,0.242) 
(0.64,0.243) 
(0.65,0.248) 
(0.66,0.248) 
(0.67,0.248) 
(0.68,0.22) 
(0.69,0.221) 
(0.7,0.221) 
(0.71,0.221) 
(0.72,0.193) 
(0.73,0.195) 
(0.74,0.167) 
(0.75,0.167) 
(0.76,0.167) 
(0.77,0.168) 
(0.78,0.169) 
(0.79,0.169) 
(0.8,0.17) 
(0.81,0.176) 
(0.82,0.176) 
(0.83,0.165) 
(0.84,0.168) 
(0.85,0.161) 
(0.86,0.161) 
(0.87,0.161) 
(0.88,0.162) 
(0.89,0.162) 
(0.9,0.162) 
(0.91,0.163) 
(0.92,0.163) 
(0.93,0.163) 
(0.94,0.163) 
(0.95,0.163) 
(0.96,0.164) 
(0.97,0.156) 
(0.98,0.157) 
(0.99,0.157) 
(1.0,0.204) 
        };
        \addlegendentry{Hard Bounding Box}
        \draw [stealth-,thick] (axis cs:0.11,0.43) -- (axis cs:0.18,0.43) node[right]{0.11} ;

    \addplot
        [color=blue]
        coordinates {
(0.0,0.055) 
(0.01,0.096) 
(0.02,0.107) 
(0.03,0.11) 
(0.04,0.136) 
(0.05,0.19) 
(0.06,0.198) 
(0.07,0.227) 
(0.08,0.294) 
(0.09,0.274) 
(0.1,0.315) 
(0.11,0.298) 
(0.12,0.341) 
(0.13,0.376) 
(0.14,0.359) 
(0.15,0.279) 
(0.16,0.282) 
(0.17,0.284) 
(0.18,0.285) 
(0.19,0.294) 
(0.2,0.295) 
(0.21,0.3) 
(0.22,0.29) 
(0.23,0.291) 
(0.24,0.304) 
(0.25,0.304) 
(0.26,0.305) 
(0.27,0.308) 
(0.28,0.31) 
(0.29,0.32) 
(0.3,0.329) 
(0.31,0.334) 
(0.32,0.35) 
(0.33,0.322) 
(0.34,0.334) 
(0.35,0.337) 
(0.36,0.335) 
(0.37,0.311) 
(0.38,0.317) 
(0.39,0.317) 
(0.4,0.316) 
(0.41,0.297) 
(0.42,0.298) 
(0.43,0.302) 
(0.44,0.316) 
(0.45,0.296) 
(0.46,0.298) 
(0.47,0.298) 
(0.48,0.298) 
(0.49,0.303) 
(0.5,0.29) 
(0.51,0.273) 
(0.52,0.274) 
(0.53,0.274) 
(0.54,0.247) 
(0.55,0.247) 
(0.56,0.248) 
(0.57,0.248) 
(0.58,0.248) 
(0.59,0.249) 
(0.6,0.249) 
(0.61,0.222) 
(0.62,0.223) 
(0.63,0.201) 
(0.64,0.201) 
(0.65,0.202) 
(0.66,0.201) 
(0.67,0.202) 
(0.68,0.202) 
(0.69,0.203) 
(0.7,0.204) 
(0.71,0.204) 
(0.72,0.205) 
(0.73,0.205) 
(0.74,0.178) 
(0.75,0.178) 
(0.76,0.179) 
(0.77,0.18) 
(0.78,0.181) 
(0.79,0.182) 
(0.8,0.182) 
(0.81,0.182) 
(0.82,0.182) 
(0.83,0.183) 
(0.84,0.162) 
(0.85,0.161) 
(0.86,0.161) 
(0.87,0.163) 
(0.88,0.163) 
(0.89,0.165) 
(0.9,0.166) 
(0.91,0.166) 
(0.92,0.152) 
(0.93,0.153) 
(0.94,0.126) 
(0.95,0.126) 
(0.96,0.126) 
(0.97,0.126) 
(0.98,0.126) 
(0.99,0.127) 
(1.0,0.177) 
        };
        \addlegendentry{Soft Bounding Box}
        \draw [stealth-,thick] (axis cs:0.13,0.376) -- (axis cs:0.13,0.2) node[below]{0.13} ;

    \end{axis}
    ;
  \end{tikzpicture}
  \caption[Ex2]%
  {{\small NNDR, Diveristy}}    
\end{subfigure}
\begin{subfigure}[b]{0.4\textwidth}
  \centering
\begin{tikzpicture}[scale=0.6]
    \begin{axis}[
        name=plot1,
        title={},
        xlabel={$\rho_{th}$},
        ylabel={$\mathrm{A-MRR}$},
        xmin=0, xmax=1,
        ymin=0, ymax=0.45,
        xtick={0,0.2,0.40,0.60,0.80,1.00},
        ytick={0,0.20,0.40,0.60,0.80,1.00},
        legend pos=north east,
        ymajorgrids=false,
        grid style=dashed,
    ]

    \addplot
        [color=red]
        coordinates {
(0.0,0.055) 
(0.01,0.204) 
(0.02,0.235) 
(0.03,0.247) 
(0.04,0.333) 
(0.05,0.363) 
(0.06,0.349) 
(0.07,0.34) 
(0.08,0.305) 
(0.09,0.269) 
(0.1,0.269) 
(0.11,0.254) 
(0.12,0.177) 
(0.13,0.178) 
(0.14,0.153) 
(0.15,0.128) 
(0.16,0.136) 
(0.17,0.142) 
(0.18,0.143) 
(0.19,0.143) 
(0.2,0.121) 
(0.21,0.124) 
(0.22,0.101) 
(0.23,0.124) 
(0.24,0.107) 
(0.25,0.11) 
(0.26,0.088) 
(0.27,0.088) 
(0.28,0.091) 
(0.29,0.103) 
(0.3,0.103) 
(0.31,0.103) 
(0.32,0.103) 
(0.33,0.103) 
(0.34,0.103) 
(0.35,0.103) 
(0.36,0.103) 
(0.37,0.103) 
(0.38,0.103) 
(0.39,0.103) 
(0.4,0.103) 
(0.41,0.103) 
(0.42,0.103) 
(0.43,0.075) 
(0.44,0.053) 
(0.45,0.053) 
(0.46,0.053) 
(0.47,0.053) 
(0.48,0.053) 
(0.49,0.053) 
(0.5,0.053) 
(0.51,0.054) 
(0.52,0.054) 
(0.53,0.055) 
(0.54,0.055) 
(0.55,0.055) 
(0.56,0.055) 
(0.57,0.055) 
(0.58,0.055) 
(0.59,0.055) 
(0.6,0.055) 
(0.61,0.055) 
(0.62,0.055) 
(0.63,0.055) 
(0.64,0.055) 
(0.65,0.055) 
(0.66,0.055) 
(0.67,0.055) 
(0.68,0.055) 
(0.69,0.055) 
(0.7,0.055) 
(0.71,0.055) 
(0.72,0.055) 
(0.73,0.055) 
(0.74,0.055) 
(0.75,0.055) 
(0.76,0.055) 
(0.77,0.055) 
(0.78,0.055) 
(0.79,0.055) 
(0.8,0.055) 
(0.81,0.055) 
(0.82,0.055) 
(0.83,0.055) 
(0.84,0.055) 
(0.85,0.055) 
(0.86,0.055) 
(0.87,0.055) 
(0.88,0.055) 
(0.89,0.055) 
(0.9,0.055) 
(0.91,0.055) 
(0.92,0.055) 
(0.93,0.055) 
(0.94,0.055) 
(0.95,0.055) 
(0.96,0.055) 
(0.97,0.055) 
(0.98,0.055) 
(0.99,0.055) 
(1.0,0.055) 
        };
        \addlegendentry{Full Image}
        \draw [stealth-,thick] (axis cs:0.05,0.363) -- (axis cs:0.15,0.32) node[right]{0.05} ;
        
        \addplot
        [color=green]
        coordinates {
(0.0,0.063) 
(0.01,0.297) 
(0.02,0.37) 
(0.03,0.343) 
(0.04,0.353) 
(0.05,0.341) 
(0.06,0.299) 
(0.07,0.294) 
(0.08,0.226) 
(0.09,0.251) 
(0.1,0.225) 
(0.11,0.215) 
(0.12,0.216) 
(0.13,0.217) 
(0.14,0.218) 
(0.15,0.22) 
(0.16,0.203) 
(0.17,0.203) 
(0.18,0.182) 
(0.19,0.156) 
(0.2,0.135) 
(0.21,0.136) 
(0.22,0.138) 
(0.23,0.138) 
(0.24,0.138) 
(0.25,0.141) 
(0.26,0.113) 
(0.27,0.113) 
(0.28,0.113) 
(0.29,0.114) 
(0.3,0.096) 
(0.31,0.096) 
(0.32,0.07) 
(0.33,0.07) 
(0.34,0.081) 
(0.35,0.081) 
(0.36,0.059) 
(0.37,0.059) 
(0.38,0.057) 
(0.39,0.057) 
(0.4,0.057) 
(0.41,0.058) 
(0.42,0.058) 
(0.43,0.058) 
(0.44,0.058) 
(0.45,0.058) 
(0.46,0.058) 
(0.47,0.058) 
(0.48,0.058) 
(0.49,0.058) 
(0.5,0.058) 
(0.51,0.058) 
(0.52,0.059) 
(0.53,0.059) 
(0.54,0.059) 
(0.55,0.059) 
(0.56,0.059) 
(0.57,0.059) 
(0.58,0.059) 
(0.59,0.059) 
(0.6,0.059) 
(0.61,0.059) 
(0.62,0.059) 
(0.63,0.059) 
(0.64,0.059) 
(0.65,0.063) 
(0.66,0.063) 
(0.67,0.063) 
(0.68,0.063) 
(0.69,0.063) 
(0.7,0.063) 
(0.71,0.063) 
(0.72,0.063) 
(0.73,0.063) 
(0.74,0.063) 
(0.75,0.063) 
(0.76,0.063) 
(0.77,0.063) 
(0.78,0.063) 
(0.79,0.063) 
(0.8,0.063) 
(0.81,0.063) 
(0.82,0.063) 
(0.83,0.063) 
(0.84,0.063) 
(0.85,0.063) 
(0.86,0.063) 
(0.87,0.063) 
(0.88,0.063) 
(0.89,0.063) 
(0.9,0.063) 
(0.91,0.063) 
(0.92,0.063) 
(0.93,0.063) 
(0.94,0.063) 
(0.95,0.063) 
(0.96,0.063) 
(0.97,0.063) 
(0.98,0.063) 
(0.99,0.063) 
(1.0,0.063) 
        };
        \addlegendentry{Hard Bounding Box}
        \draw [stealth-,thick] (axis cs:0.02,0.370) -- (axis cs:0.15,0.37) node[right]{0.02} ;

    \addplot
        [color=blue]
        coordinates {
(0.0,0.055) 
(0.01,0.189) 
(0.02,0.205) 
(0.03,0.277) 
(0.04,0.421) 
(0.05,0.414) 
(0.06,0.4) 
(0.07,0.376) 
(0.08,0.314) 
(0.09,0.317) 
(0.1,0.28) 
(0.11,0.226) 
(0.12,0.227) 
(0.13,0.201) 
(0.14,0.159) 
(0.15,0.137) 
(0.16,0.14) 
(0.17,0.141) 
(0.18,0.143) 
(0.19,0.166) 
(0.2,0.169) 
(0.21,0.174) 
(0.22,0.178) 
(0.23,0.178) 
(0.24,0.158) 
(0.25,0.158) 
(0.26,0.131) 
(0.27,0.133) 
(0.28,0.111) 
(0.29,0.11) 
(0.3,0.11) 
(0.31,0.103) 
(0.32,0.103) 
(0.33,0.103) 
(0.34,0.103) 
(0.35,0.103) 
(0.36,0.103) 
(0.37,0.103) 
(0.38,0.103) 
(0.39,0.103) 
(0.4,0.103) 
(0.41,0.081) 
(0.42,0.081) 
(0.43,0.081) 
(0.44,0.081) 
(0.45,0.081) 
(0.46,0.081) 
(0.47,0.081) 
(0.48,0.053) 
(0.49,0.053) 
(0.5,0.054) 
(0.51,0.054) 
(0.52,0.054) 
(0.53,0.054) 
(0.54,0.054) 
(0.55,0.054) 
(0.56,0.054) 
(0.57,0.054) 
(0.58,0.054) 
(0.59,0.055) 
(0.6,0.055) 
(0.61,0.055) 
(0.62,0.055) 
(0.63,0.055) 
(0.64,0.055) 
(0.65,0.055) 
(0.66,0.055) 
(0.67,0.055) 
(0.68,0.055) 
(0.69,0.055) 
(0.7,0.055) 
(0.71,0.055) 
(0.72,0.055) 
(0.73,0.055) 
(0.74,0.055) 
(0.75,0.055) 
(0.76,0.055) 
(0.77,0.055) 
(0.78,0.055) 
(0.79,0.055) 
(0.8,0.055) 
(0.81,0.055) 
(0.82,0.055) 
(0.83,0.055) 
(0.84,0.055) 
(0.85,0.055) 
(0.86,0.055) 
(0.87,0.055) 
(0.88,0.055) 
(0.89,0.055) 
(0.9,0.055) 
(0.91,0.055) 
(0.92,0.055) 
(0.93,0.055) 
(0.94,0.055) 
(0.95,0.055) 
(0.96,0.055) 
(0.97,0.055) 
(0.98,0.055) 
(0.99,0.055) 
(1.0,0.055) 
        };
        \addlegendentry{Soft Bounding Box}
        \draw [stealth-,thick] (axis cs:0.04,0.421) -- (axis cs:0.15,0.421) node[right]{0.04} ;

    \end{axis}
    ;
  \end{tikzpicture}
  \caption[Ex2]%
  {{\small TVSS, Diveristy}}    
\end{subfigure}
\caption[ Training the thresholds ]
{\small Training the thresholds using saliency maps for $g$.} 
\label{fig:thresholds}
\end{figure*}

%% file: 4_0_experiments.tex
\section{Experiments}
\label{sec:experiments}


The proposed system was trained and evaluated in a subset of images from the NTCIR-Lifelog dataset \cite{NTCIR-Lifelog} according to the Mean Reciprocal Rank (MRR) \cite{Craswell2009}.
The details and results are presented and discussed in this section.


\input{4_1_dataset}

\input{4_2_metric}

\input{4_3_training}

\input{4_4_test}

%% file: 4_1_dataset.tex
\subsection{Datasets}
\label{sec:dataset}

Our experiments used the NII Testbeds and Community for Information access Research (NTCIR) Lifelog dataset, which is composed of a total of 88,185 images acquired by 3 people using an Autographer camera during 90 days, 30 days per person. In our experiments, though, we only considered one of the users. 
The Autographer camera used in the NTCIR-Lifelog dataset uses a wide angle lens, a feature which resulted helpful as the images where more likely to include personal objects. 

\subsubsection{Definition of Queries}
\label{subsec:queries}
When performing a search the system needs an input of some images of the object in order to look for it. To carry out the experiments, and after doing an exhaustive analysis of the dataset, we decided to work with four object categories: \textit{mobile phone}, \textit{laptop}, \textit{watch}, and \textit{headphones}. 
The set $Q$ was built containing five images of the own dataset for each category. The whole object was present in these images and occupied most of it.

The query images were manually annotated with a bounding box to assess the \textit{Hard Bounding Box (HBB)} and \textit{Soft Bounding Box (SBB)} configurations. In our dataset, this corresponded to a total of 25 bounding boxes. We consider that this scenario is realistic because, if necessary, a user could be ask to manually annotate five images of the object he is looking for. However, the results presented later in Section \ref{sec:test} show that this task may not be even necessary.


\begin{figure}[h!]
\centering
\includegraphics[width=\columnwidth]{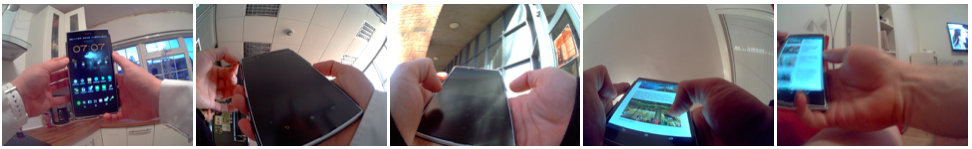}
\caption[Images to build the query]{Set of images $Q$ used to build the query vector $\vec{q}$ for the category \textit{mobile phone}.}
\label{fig:queries}
\end{figure}


\subsubsection{Annotation of the Dataset}
We decided to build the annotations of the dataset following this guideline: \textit{"Consider as relevant those images that would help to find out where was the last time that the camera saw the object"}. 
This strategy made us considering as relevant all the images that depicted both the location and the object. 
Any of them would help the user to find his object. 

%% file: 4_2_metric.tex
\subsection{Evaluation metric}
\label{sec:metric}

To assess the performance of the system, an evaluation metric must be chosen to be able to quantitatively compare how different configurations perform.
This metric should be as realistic as possible and must have the ability to measure exactly whether or not the system helps the user when he or she looks for the objects.









The Mean Reciprocal Rank ($\mathrm{MRR}$)~\cite{Craswell2009} is the average of the reciprocal ranks of results for a sample of queries $Q$, being the reciprocal rank of a query response the inverse in the rank of the first relevant answer $q^*$. For given a day $d$, its mathematical expression is:

\begin{equation}
\mathrm{MRR_d} = \frac{1}{|Q_d|}\sum_{q \in Q_d}\frac{1}{q^*}
\label{eq:mrr}
\end{equation}

Mean Reciprocal Rank is associated with a user model where the user only wishes to see one relevant document.
We have defined the Averaged-MRR (AMRR) to refer to the average of $\mathrm{MRR}$s obtained across test all days. Given a set of days $D = \left \lbrace d_1,d_2,...,d_k\right\rbrace$ the expression of the Averaged Mean Reciprocal Rank is

\begin{equation}
\mathrm{AMRR} = \frac{1}{|D|}\sum_{d \in D}\mathrm{MRR_d} = \frac{1}{|D|}\sum_{d \in D}\frac{1}{|Q_d|}\sum_{q \in Q_d}\frac{1}{q^*}
\label{eq:amrr}
\end{equation}


%% file: 4_3_training.tex
\subsection{Training}
\label{sec:training}




The 30 days of data available for one user were used in the following way:  
the queries were defined using 3 days, the training partition included 9 different days, and the testing was performed on the remaining 15 days. The remaining 3 days were discarded as they did not include enough quality appearances of the objects to be considered.


The values that we wanted to train were the construction of the codebook for the visual words as well as the thresholds used in both techniques described in \ref{sec:candidate}, TVSS and NNDR.

\begin{itemize}

\item \textit{Visual Words Codebook}: The BoW framework defined by \cite{eva2016bags} requires building a visual codebook in order to map vectors to their nearest centroid. This codebook was built using $k$-means clustering.
We used an accelerated algorithm based on approximate nearest neighbors on local CNN features to fit a codebook with 25,000 centroids as in \cite{eva2016bags}.

\item \textit{Threshold values $\nu_{th}$ and $\rho_{th}$}: To predict what would be the best value for these parameters, the same procedure was applied for both. We performed a sweep from 0 to 1 with a step-size of 0.01. For each of these values the MRR was computed and averaged across the 9 days that composed the training set. Therefore, this AMRR can be understood as a function of the threshold, so an optimal argument can be chosen. Figure~\ref{fig:thresholds} shows the curves obtained and the optimal values chosen when training with saliency maps for the $g$ function. When training using other configurations for the $g$ function, the optimal thresholds $\rho_{th}$ and $\nu_{th}$ remained in the same values despite AMRR being slightly different. 

\end{itemize}

%% file: 4_4_test.tex
\subsection{Testing}
\label{sec:test}


The different methods presented in Section \ref{sec:methodology} are assessed in this section over a test set of 15 days from the NTCIR Lifelog dataset.
Table~\ref{tab:parameterssummary} summarizes the configuration options for the different stages of the pipeline. 
The study aims at identifying which is the impact of each of the proposed methods in the complete solution.


\begin{table}[ht!]
\centering
\begin{tabular}{@{}clc@{}}
\toprule
Method                                                                                       & \multicolumn{1}{c}{} & Options     \\ \midrule
\multirow{3}{*}{\begin{tabular}[c]{@{}c@{}}Query \end{tabular}} & \multicolumn{1}{c}{} & Full Image (FI)         \\
                                                                                           &                      & Hard Bounding Box (HBB) \\
                                                                                           &                      & Soft Bounding Box (SBB) \\ \midrule
\multirow{3}{*}{\begin{tabular}[c]{@{}c@{}}Target database\end{tabular}}    & \multicolumn{1}{c}{} & Full Image (FI)         \\
                                                                                           &                      & Center Bias (CB)      \\
                                                                                           &                      & Saliency Map (SM)     \\ \midrule
\multirow{2}{*}{Thresholding criterion}                                                         & \multicolumn{1}{c}{} & Nearest Neighbors (NNDR)                    \\
                                                                                           &                      & Similarity Score (TVSS)                    \\ \midrule
\multirow{2}{*}{Ranking}                                                  &                      & Time Sorting      \\
                                                                                           & \multicolumn{1}{c}{} & Interleaving (I)           \\ \bottomrule
\end{tabular}
\caption{Configuration parameters summary.}
\label{tab:parameterssummary}
\end{table}

Tables~\ref{tab:fullimage}, \ref{tab:weighted} and \ref{tab:saliency} contain the  AMRR values obtained for all possible configurations, being \textit{Time Sorting} the baseline.  Visual Ranking is included to understand the impact of the proposed temporal reranking stages with respect to the visual search obtained with \cite{eva2016bags}.

\input{tables}


Comparing the results obtained with any of the system configurations (the four last columns) versus the intermediate stage (the visual ranking $R_v$) and the defined baseline (the temporal sorting), we can conclude that all proposed methods both the visual search and the diversity-based reranking improve the AMRR. In other words, all configurations provide a faster option to find the last appearance of the lost object than brute-force approach of just browsing backwards in time through the sequence of egocentric images.

Comparing each approach with or without temporal diversity indicates a gain in all cases but in the SBB-NNDR one of Table~\ref{tab:fullimage}, where performance slightly drops. Our study reveals that, in general, a temporal interleaving of the results is a good choice. This is true in all the best configurations of Tables~\ref{tab:fullimage}, \ref{tab:weighted} and~\ref{tab:saliency}.

A comparison between Table~\ref{tab:fullimage} and~\ref{tab:weighted} shows that weighting the convolutional features of the target dataset with a a central bias does not improve the results obtained when using the full image query, despite this strategy has been used in other works related to salient object detection~\cite{liu2011learning}.
We suspect that this is due objects in egocentric images not being located in the center of the image as often as they are for intentionally taken photographs.
This same conclusion was reached in~\cite{buso2015geometrical}.

Focusing in Table \ref{tab:saliency} indicates that weighting the convolutional features of the target images with a saliency map is only beneficial when the full query image is considered, actually providing the best results among all configurations.
However, saliency maps decrease the performance when the a hard bounding box defines the query, and does not introduce much changes when a soft bounding box is considered.
In other words, focusing on the local information of the object in the query image may not be beneficial, and exploiting its the context may help.
Notice though that using the full query image is only the best of the three query configurations when the convolutional features of the target images are weighted by the saliency maps.
We hypothesize that the saliency maps, apart from identifying the local features of the object in the target image, it also emphasizes other features in the background that boost the matching with the full query images.
In these later cases, we would be exploiting the case when an object tends to appear in the same locations, somehow providing a prior for our system.
When an object is lost, looking at the places where we have use it frequently in the past may be beneficial, and the features characterizing the location are found outside the local query mask.




%% file: tables.tex
\begin{table*}[h!]
\centering
\begin{tabular}{@{}ccccccc@{}}
\toprule
$f(Q)$              & Time Sorting & Visual Ranking & NNDR  & TVSS  & NNDR+I~\protect\footnotemark & TVSS+I \\ \midrule
FI       &                   & 0,157          & 0,216 & 0,213 & 0,231    & 0,223    \\
HBB         & 0,051             & 0,139          & 0,212 & 0,180 & 0,216    & 0,184    \\
SBB &                   & 0,163          & 0,171 & 0,257 & 0,169    & \textbf{0,269}    \\ \bottomrule
\end{tabular}
\caption[Results using Full Image for $g$]{$\mathrm{A \operatorname{-} MRR}$ using Full Image for $g$.}
\label{tab:fullimage}
\end{table*}

\footnotetext{I stands for Interleaving}


\begin{table*}[h!]
\centering
\begin{tabular}{@{}ccccccc@{}}
\toprule
$f(Q)$              & Time Sorting & Visual Ranking & NNDR  & TVSS  & NNDR+I & TVSS+I \\ \midrule
FI       &                   & 0,156          & 0,191 & 0,205 & 0,206    & 0,215    \\
HBB          & 0,051             & 0,130          & 0,212 & 0,170 & 0,216    & 0,174    \\
SBB &                   & 0,162          & 0,160 & 0,240 & 0,161    & \textbf{0,258}    \\ \bottomrule
\end{tabular}
\caption[Results using Center Bias for $g$]{$\mathrm{A \operatorname{-} MRR}$ using Center Bias for $g$.}
\label{tab:weighted}
\end{table*}


\begin{table*}[h!]
\centering
\begin{tabular}{@{}ccccccc@{}}
\toprule
$f(Q)$              & Time Sorting & Visual Ranking & NNDR  & TVSS  & NNDR+I & TVSS+I \\ \midrule
FI       &                   & 0,150          & 0,240 & 0,274 & 0,249    & \textbf{0,283}    \\
HBB          & 0,051             & 0,173          & 0,200 & 0,136 & 0,206    & 0,147    \\
SBB &                   & 0,178          & 0,168 & 0,242 & 0,174    & 0,257    \\ \bottomrule
\end{tabular}
\caption[Results using Saliency Maps for $g$]{$\mathrm{A \operatorname{-} MRR}$ using Saliency Maps for $g$.}

\label{tab:saliency}
\end{table*}

%% file: 5_0_conclusions.tex
\section{Conclusions}
\label{sec:conclusions}

The main objective of this work was to design a retrieval system to find personal objects in egocentric images.
Compared to the proposed baseline, the contributions reported in this document have shown that the system is helpful for the task.
We believe these results might be useful as a baseline for further research on this field. 


An interesting observation of this work is the fact that, despite being a common strategy in many other tasks, applying a center bias weighting did not improve results, but weighting with saliency maps improved performance significantly.



As a future work, we suggest to explore different approaches for the temporal reordering stage that might improve the system performance. Referring to the visual part, fine-tuning could be performed to adapt the network to the egocentric images and improve its improve the accuracy when extracting the local convolutional features.

%% file: acknowledgements.tex
\section{Acknowledgments}


The main author developed this work thanks to the financial support of the Erasmus+ Programme for Student Mobility in the European Union, Generalitat de Catalunya and Centre de Formaci\'o Interdisciplinaria Superior (CFIS).
This publication has emanated from research conducted with the financial support of Science Foundation Ireland (SFI) under grant number SFI/12/RC/2289.
This work has been developed in the framework of the project BigGraph TEC2013-43935-R, funded by the Spanish Ministerio de Econom'ia y Competitividad and the European Regional Development Fund (ERDF). 
The Image Processing Group at the UPC is a SGR14 Consolidated Research Group recognized and sponsored by the Catalan Government (Generalitat de Catalunya) through its AGAUR office. 
We gratefully acknowledge the support of NVIDIA Corporation with the donation of the GeForce GTX Titan Z used in this work.

